\documentclass{article}
\usepackage{spconf,amsfonts,latexsym,amssymb,amsmath,epsfig}

\newcommand{\bx}{\mathbf{x}}
\newcommand{\bX}{\mathbf{X}}
\newcommand{\bbC}{\mathbb{C}}

\newcommand{\hgamma}{\hat{\gamma}}
\newcommand{\cE}{\mathcal{E}}
\newcommand{\cov}{\operatorname{cov}}
\newcommand{\ha}{\hat{a}}
\newcommand{\hb}{\hat{b}}

\title{MAXIMUM-ENTROPY SURROGATION IN NETWORK SIGNAL DETECTION}

\name{D. Cochran$^1$, S. D. Howard\,$^2$, B. Moran$\,^3$, H. A. Schmitt$\,^4$
\thanks{This work was supported in part by the University of Michigan and the U.S.~Army 
        Research Office under MURI award No.~W911NF-11-1-0391 and by Physical Layer Systems, LLC.}}
\address{$^1$Arizona State University, Tempe AZ,  USA\\
         $^2$Defence Science \& Technology Organisation, Edinburgh SA, Australia\\
         $^3$Defence Science Institute, Parkville VIC, Australia \\
         $^4$Physical Layer Systems LLC, Tucson AZ, USA}

\begin{document}
\maketitle
\begin{abstract}
Multiple-channel detection is considered in the context of a sensor network where raw data are shared only by nodes that have a common edge in the network graph.  Established multiple-channel detectors, such as those based on generalized coherence or multiple coherence, use pairwise measurements from every pair of sensors in the network and are thus directly applicable only to networks whose graphs are completely connected.  An approach introduced here uses a maximum-entropy technique to formulate surrogate values for missing measurements corresponding to pairs of nodes that do not share an edge in the network graph. The broader potential merit of maximum-entropy baselines in quantifying the value of information in sensor network applications is also noted. 
\end{abstract}
\section{Introduction}
\label{sec:intro}

Established methods in coherent multiple-channel signal detection typically assume that data from all sensors is collected at a single location for processing. In particular, this view is implicit in the generalized coherence (GC) approach introduced in \cite{Gish87} and elaborated and extended in numerous other works (e.g., \cite {Cochran95,Clausen97,Clausen01,Ramirez10,Songsri12}). In signal processing for sensor networks, it is frequently desirable to process data locally at (or in local neighborhoods of) the nodes and reduce, or even eliminate, the need for aggregation of data at a ``fusion center.''

In GC-based detection and related methods, such as those using multiple coherence \cite{Trueblood78}, processing entails computing inner products (correlations) between segments of time series data collected at each pair of nodes in the network.  When all data are collected at a fusion center, this is not an issue. This paper proposes an approach for implementing a GC detector that incorporates a high degree of local data reduction by using only inner products of time series segments collected at pairs of nodes that are adjacent in the topology of the network.  This is equivalent to traditional GC detection only when the network is fully connected.  Otherwise, the detector must operate without using the inner products associated with pairs of nodes that do not share an edge in the network graph.

The approach introduced here replaces each missing inner product datum by a surrogate value obtained via a maximum entropy method, and then proceeds to apply a standard GC detector as though all data were available.  The paper begins in Section \ref{sec:GC_background} by summarizing the basics of GC detection, and proceeds in Section \ref{sec:max_ent} to describe the maximum entropy procedure that allows surrogation for missing data in networks that are not fully connected.  Section \ref{sec:results} shows simulation results for small networks that illustrate the performance of the approach in such settings.  The paper concludes in Section \ref{sec:conclusion} with a discussion of further work needed to make this technique viable for larger sensor networks and also how this approach suggests a point of view about quantifying the value of information in network signal processing.

\section{The generalized coherence detector}
\label{sec:GC_background}

The GC estimate is an established statistic for detection of a common signal on several noisy channels \cite{Cochran95}.  Its properties and applications have been well documented, and it continues to be extended and studied in various contexts \cite{Ramirez10,Songsri12}. A crucial drawback of GC-based methods in the sensor network setting is that standard implementations require all the raw sensor data to be collected in one place (i.e., a ``fusion center'') to perform the processing. 

Given $M$ measurements $X_1,\ldots,X_M$ at the nodes of a network with each $X_m\in\bbC^N$, define normalized measurement vectors $U_m=X_m/||X_m||$ for $m=1,\ldots,M$.  The GC estimate obtained from these measurements is $\hgamma^2(X_1,\ldots,X_M)=1-\det G(U_1,\ldots,U_M)$, where the matrix $G(U_1,\ldots,U_M)$ is the $M\times M$ Gram matrix
\begin{equation}
\label{eq:normalized_Gram}
\left[ \begin{array}{cccc} 
1 & \left<U_1,U_2\right> & \cdots & \left<U_1,U_M\right>\\
\left<U_2,U_1\right> & 1 & \cdots & \left<U_2,U_M\right>\\
\vdots & & \ddots & \vdots \\
\left<U_M,U_1\right> & \cdots & \left<U_M,U_{M-1}\right> & 1
\end{array}\right]
\end{equation}
formed from the normalized data vectors. In typical multiple-channel detection applications, the value of $\hgamma^2$ is compared to a threshold to decide between signal-present ($H_1$) and signal-absent ($H_0$) hypotheses.

\begin{figure}[hbt]
\begin{center}
\includegraphics[width=0.75\linewidth]{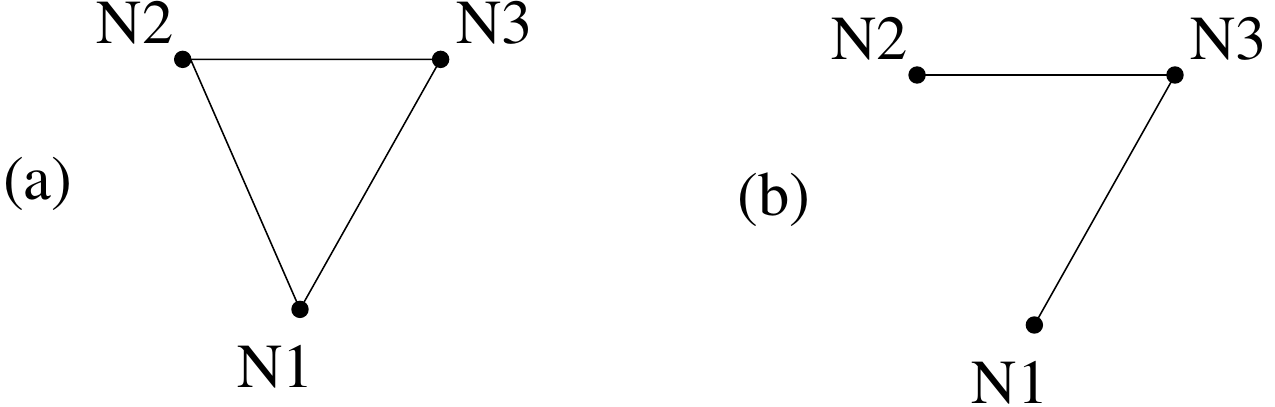}
\end{center}
\caption{(a) A completely connected three-node network. (b)
         A three-node network with no edge between nodes one and two. 
         Every three-node network with two edges has the same 
         topology (i.e., linear).}
\label{fig:3node}
\end{figure}

In a completely connected network, such as those depicted in Figures \ref{fig:3node}(a) and \ref{fig:4node}(a), all of the inner products comprising the Gram matrix may be computed locally; i.e., by exchange of data between nodes that share an edge in the network graph. This enables dramatic reduction in the amount of data necessary to communicate to the fusion center in order to implement a GC detector for the network.  When the network graph is not complete, however, inner products of data vectors corresponding to nodes that do not share an edge cannot be computed locally and transmission of these scalar values corresponding to each edge in the network graph is not sufficient to enable the fusion center to compute the GC estimate.  

Considering the three-node case depicted in Figure \ref{fig:3node}(b), note that the value of $\left<U_1,U_2\right>$ is generally not determined by the values of $\left<U_1,U_3\right>$ and $\left<U_2,U_3\right>$. But it is not independent of these values either; {\em e.g.}, $G$ must be non-negative definite. In the approach described in Sec.~\ref{sec:max_ent}, the values of the ``missing'' inner products ({\em i.e.}, those corresponding to pairs of nodes that do not share an edge) will be replaced by surrogate values obtained via a maximum entropy technique, thereby enabling a GC test to be performed despite the missing data.  Sec.~\ref{sec:results} considers the detection performance of GC detectors implemented in this way.

\section{Maximum-entropy surrogation}
\label{sec:max_ent}

Assume that there is a complex random variable $\bx_m$ associated with each network node $m$, modeling data samples collected at that node. Collected samples at node $m$ are hence realizations of $\bx_m$ and can be used to estimated the mean and variance of $\bx_m$ in standard ways \cite{Papoulis02}. The ability to communicate between nodes $m$ and $j$ linked by an edge permits estimation of the covariance $\cov(\bx_m,\bx_j)$ by similar methods. For a complete graph ({\em i.e.}, one in which each pair of nodes shares an edge), it is thus possible to estimate the full $M\times M$ covariance matrix $C$ of the variables $\bx_1,\ldots,\bx_M$. Indeed, assuming the $X_m$ have mean zero, if $N$ independent samples of $\bx_m$ are collected at node $m$ for each $m$ and are assimilated into complex $N$-vectors $\bX_1,\ldots,\bX_M$ of sample values, then the standard estimator $\widehat{C}$ of $C$ is proportional to the Gram matrix $G(\bX_1,\ldots,\bX_M)$. The GC detector can thus be viewed as a test on the estimated covariance matrix of the variates $\bx_1,\ldots,\bx_M$. If each $\bx_m$ is further assumed to be normalized to unit variance, using the true variance in place of an estimate on the main diagonal gives a test matrix of the form (\ref{eq:normalized_Gram}).

The maximum-entropy method \cite{jaynes79} holds that missing values in $C$ should be surrogated in such a way as to introduce no new assumptions about the nature of the random variables or of the network. The joint distribution of the random variables $\bx_1,\ldots,\bx_M$ that best describes current knowledge ({\em i.e.}, the covariance estimates for all pairs of directly connected nodes) with no further assumptions is the maximum entropy distribution constrained by the available data.

The problem of finding the maximum-entropy completion of a covariance matrix has been studied in prior literature (see, {\em e.g.}, \cite{Vandenberghe98} and references cited therein). The maximum-entropy probability density $p(x_1,\ldots,x_M)$ consistent with the estimated covariances must be of the form
\begin{equation}
  \label{eq:maxent}
  \frac{1}{Z}\exp\left\{ -\Bigl(\sum_{m=1}^M \lambda_m x_m^*+\frac{1}{2}\sum_{(m,j)\in\cE}\mu_{m,j}x_m x_j\Bigr)\right\}
\end{equation}
In this expression $\lambda_m$ for $m=1,\ldots,M$ and $\mu_{m,j}=\mu_{j,m}$ for values of $(m,j)$ corresponding to the edge set $\cE$ of the network graph are Lagrange parameters in the constrained optimization problem that arises in maximization of entropy subject to the constraints imposed by knowledge of the covariance values estimated from available measurements. For complex random variables, this will be a complex normal density and hence completely specified by its mean and covariance matrix.

It follows from (\ref{eq:maxent}), and is also noted in past literature \cite[Sec.~2.2]{Vandenberghe98}, that the covariance matrix $A$ of this maximum-entropy distribution will have the property that its inverse will has zeros in positions corresponding to the missing covariance values.  This observation gives a direct means for calculating the needed surrogate values in small networks.

{\bf Example}: Consider the three node network depicted in Fig.~\ref{fig:3node}(b). Writing the estimated covariance matrix as
\begin{equation*}
\hat{C}=\left[ \begin{array}{ccc} 1 & s & \ha \\ s^* & 1 & \hb \\ 
\ha^* & \hb^* & 1 \end{array}\right]
\end{equation*}
the value of the covariance estimate $\ha$ is assumed to be obtained by exchange of data between nodes 1 and 3 and the value $\hb$ is obtained by exchange of data between nodes 2 and 3.  A maximum-entropy surrogate value for $s$ is obtained by noting
\begin{equation*}
\hat{C}^{-1}=\frac{1}{D}
\left[ \begin{array}{ccc} 1-|\hb|^2 & \ha\hb^*-s & s\hb-\ha \\ 
\ha^* \hb - s^* & 1-|\ha|^2 & s^*\ha-\hb \\ 
s^*\hb^*-\ha^* & s\ha^*-\hb^* & 1-|s|^2 \end{array}\right]
\end{equation*}
where $D=\det\,\hat{C}$.  The 1-2 entry of $\hat{C}^{-1}$ will be zero when $s$ assumes the desired value.  Hence, $s=\ha\hb^*$ and the maximum-entropy completion of $\hat{C}$ is
\begin{equation*}
\hat{C}=\left[ \begin{array}{ccc} 1 & \ha\hb^* & \ha \\ 
\ha^*\hb & 1 & \hb \\ 
\ha^* & \hb^* & 1 \end{array}\right]
\end{equation*}

This direct calculation method was used to find surrogate values in the small network examples discussed above and evaluated in Sec.~\ref{sec:results}. In general, a necessary and sufficient condition for the existence of the maximum entropy distribution as a (normalizable) probability distribution is $\hat{C}$ is invertible as a symbolic matrix. Finding $k$ surrogate values in this way requires solving $k$ equations in $k$ unknowns, which becomes prohibitively cumbersome for even small networks ({\em e.g., $M>5$}).  Fortunately, maximum-entropy covariance matrix completion problems \cite{Gohberg95} fall into a class of determinant maximization problems (entropy in this setting is $\log \det\, \hat{C}$) that can be efficiently solved by convex programming techniques \cite{Boyd04,Vandenberghe98}.

\begin{figure}[hbt]
\begin{center}
\includegraphics[width=0.75\linewidth]{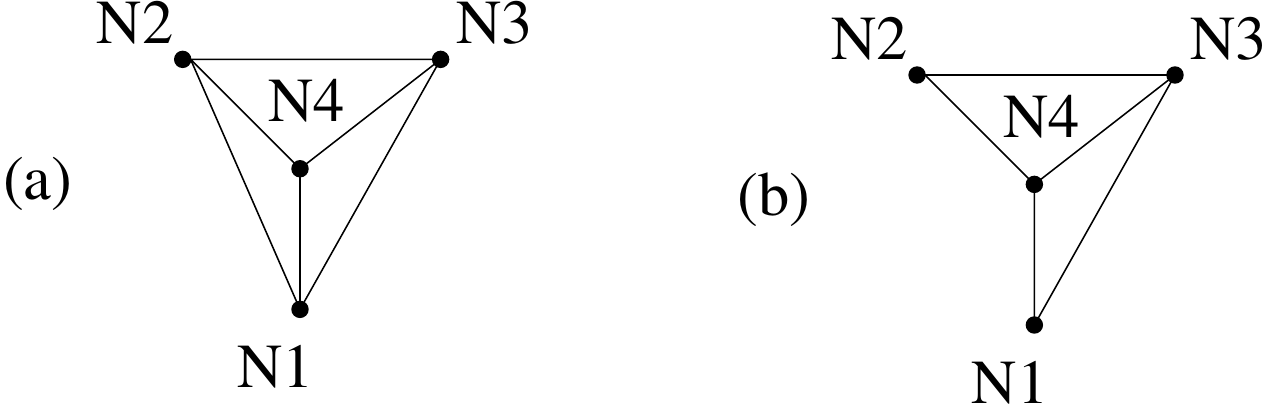}
\end{center}
\caption{(a) A completely connected four-node network. (b)
         A four-node network with no edge between nodes one and two. 
         Every four-node network with five edges has the same 
         topology.}
\label{fig:4node}
\end{figure}

\begin{figure}[hbt]
\begin{center}
\includegraphics[width=0.75\linewidth]{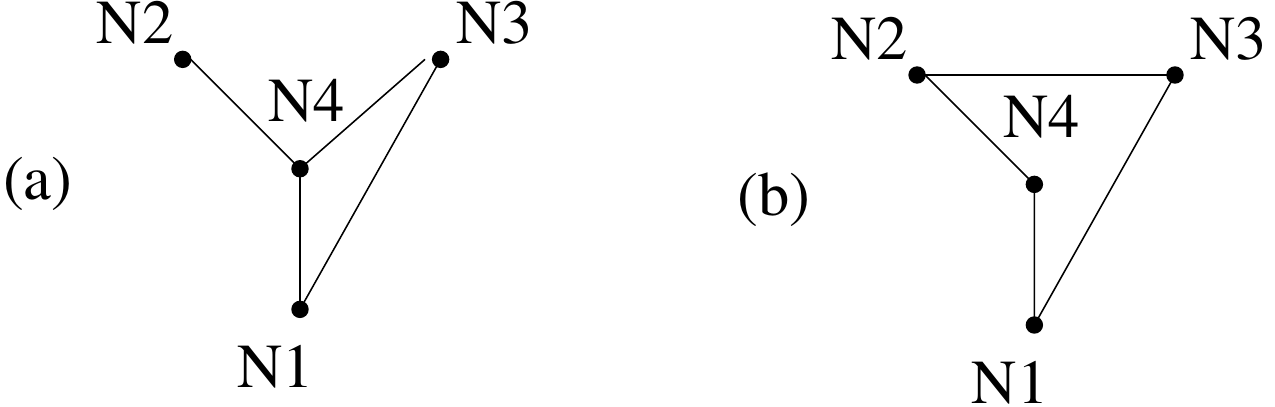}
\end{center}
\caption{Two four-node networks with no edge between nodes one and two. 
         (a) This network also lacks a 2-3 edge.
         (b) This one lacks edge 3-4 as well as 1-2.         
         Every connected four-node network with four edges has one 
         of these two topologies.}
\label{fig:4node_2}
\end{figure}

The GC statistic in invariant to re-indexing of the $M$ channels \cite{Cochran95}. This is in contrast to the multiple coherence statistic \cite{Trueblood78}, which distinguishes a reference channel. Consequently, networks that are topologically isomorphic will have identical detection performance characteristics under the approach introduced here.  For example, for a network having four nodes and five edges, it suffices to analyze the case depicted in Fig.~\ref{fig:4node}(b) since all other cases are topologically equivalent to this one (and similarly for all $M$-node networks whose graphs are one edge short of being complete).  Fig.~\ref{fig:4node_2} shows that there are topologically distinct cases of four-node networks with four edges.  In general, it will be necessary to analyze the performance of each such equivalence class separately, and it is anticipated that different topologies will provide distinct detection performance characteristics.  While further study of this aspect of the approach is beyond the scope of this paper, it is expected that comparison of network topologies based on the detection performance they support when completed via a maximum-entropy method will provide insight into sensor network design and optimization.
 
\section{Results for small networks of sensors}
\label{sec:results}

\begin{figure}[hbt]
\begin{center}
\vspace*{-2.5cm}
\includegraphics[width=1.1\linewidth]{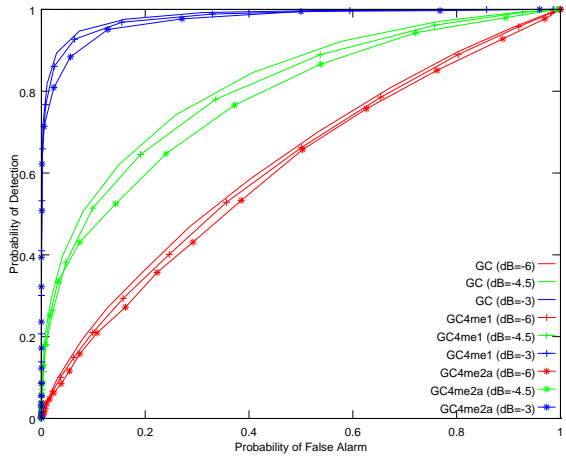}
\vspace*{-4.0cm}
\end{center}
\caption{ROC curves for detection of a 64-component white complex Gaussian signal vector in white complex Gaussian noise in a 4-node sensor network with identical SNR at each sensor. The top set of curves is at -3 dB, the center set at -4.5 dB, and the bottom set at -6 dB.  Within each set, the top curve is for the complete network (Figure \ref{fig:4node}(a)), the middle curve for the network with no link between nodes 1 and 2 (Figure \ref{fig:4node}(b)), and the bottom curve for the network with no links between nodes 1 and 2 or nodes 2 and 3 (Figure \ref{fig:4node_2}(a)).}
\label{fig:4node_results}
\end{figure}

Figure \ref{fig:4node_results} shows receiver operating characteristic (ROC) curves for detection of a white complex Gaussian signal vector in white complex Gaussian noise in a 4-node sensor network.  The vectors are of length N=64. The signal-to-noise ratio is identical at each sensor, with the top set of curves at -3 dB, the center set at -4.5 dB, and the bottom set at -6 dB.  Within each set, the top curve is for the complete network (Figure \ref{fig:4node}(a)), the middle curve for the network with no link between nodes 1 and 2 (Figure \ref{fig:4node}(b)), and the bottom curve for the network with no links between nodes 1 and 2 or nodes 2 and 3 (Figure \ref{fig:4node_2}(a)).

These curves indicate that the equal-channel SNR detection performance lost  when network connectivity is reduced by the removal of one or two links is modest -- much less significant than the detection performance would be diminished by 1 dB of SNR at each node.  While this experimental finding with such a small network is only relevant to a small application regime, it indicates that further study with larger networks is warranted.

\section{Discussion and conclusions}
\label{sec:conclusion}

A maximum entropy method to enable the implementation of generalized coherence detectors on incompletely connected sensor networks without aggregation of raw data at a fusion center has been described and demonstrated with small networks.  Performance degradation in the cases studied was modest, though more compete evaluation of the method with larger and sparser networks, unequal signal-to-noise ratios at the sensor nodes, and signals of rank greater than one is essential to establish a comprehensive understanding of this approach in detection of signals spread across a sensor network.  It is anticipated that known methods for covariance matrix completion by optimization algorithms will be essential for applying the approach to sensor networks of even moderate size ({\em e.g.}, more than ten nodes) because direct calculation of maximum-entropy surrogate values becomes formidable even for a small number of nodes.

It is noteworthy that the use of maximum entropy baselines in this application provides a mechanism for quantifying the value of information sharing within the sensor network; i.e., in this setting, a link is precisely as valuable as the performance gain it enables over the use of a maximum entropy surrogate in place of its datum.

\bibliographystyle{IEEEbib}
\bibliography{maxentGC}

\end{document}